\documentclass[aps,groupedaddress,prb,twocolumn]{revtex4}
\usepackage{graphicx,color}
\usepackage{amsmath,amssymb,amsfonts}
\usepackage{natbib}

\newcommand{\be}{\begin{equation}}
\newcommand{\ee}{\end{equation}}

\begin{document}

\title{Quantum phase analysis with quantum trajectories:\\
A step towards the creation of a Bohmian thinking}

\author{A. S. Sanz}
\affiliation{Instituto de F\'{\i}sica Fundamental -- CSIC,
Serrano 123, 28006 Madrid, Spain}

\author{S. Miret-Art\'es}
\affiliation{Instituto de F\'{\i}sica Fundamental -- CSIC,
Serrano 123, 28006 Madrid, Spain}

\date{\today}

\begin{abstract}
We introduce a pedagogical discussion on Bohmian mechanics and its
physical implications in connection with the important role played
by the quantum phase in the dynamics of quantum processes.
In particular, we have focused on phenomena such as quantum coherence,
diffraction and interference due to their historical relevance in the
development of the quantum theory and their key role in a myriad of
fundamental and applied problems of current interest.
\end{abstract}

\maketitle


\section{Introduction}
\label{sec1}

The wave function provides us with the most complete information about
a quantum system:
a probabilistic or statistical type of information about the
possible outcomes that may result after performing a measurement on
such a system.\cite{ballentine-bk}
At present, this is the generally accepted point of view,
although it has not been exempt from controversy since the very
inception of the quantum theory.
This has led to an exciting, ongoing debate on two fundamental
questions, namely the completeness of the wave function and the quantum
measurement,\cite{zurek-bk} both linked to the purpose of discerning
whether the quantum world is inherently probabilistic.
According to von Neumann's theorem,\cite{neumann-bk} this seems to be
indeed the case: quantum mechanics or any other alternative theory
cannot be formulated by simply considering a statistical approximation
from a classical-like deterministic theory.
That is, there are no more complete theories based on hidden-variable
models that can provide us with a description of quantum phenomena in
terms of dispersion-free ensembles.
Consequently, this leaves the door open for interpretations like the
observer's subjective action on quantum systems during the measurement
process, for example.

In an attempt to find an objective description of quantum phenomena,
in 1952 Bohm proposed\cite{bohm,bohm-hiley-bk,holland-bk,duerr-bk} a
physical hidden-variable model which reproduced the predictions of the
standard quantum theory without violating any of its postulates, thus
providing a counterexample to von Neumann's theorem.
Bohm's model, nowadays widely known as Bohmian mechanics,\cite{note1}
relies on the assumption that a quantum system consists of a wave and
a particle.
The wave evolves according to Schr\"odinger's equation and the particle
according to a guidance condition that makes the particle motion to be
dependent on the wave evolution.
Bohmian mechanics thus entails the very appealing feature that it
allows us to understand quantum systems on similar grounds as classical
ones, i.e., in terms of the evolution of a swarm of trajectories
through the system's configuration space.
Alternatively, this can also be interpreted as the evolution of a
quantum flow, as suggested by Madelung\cite{madelung} in 1926, thus
giving rise to a hydrodynamic form of quantum mechanics that can be
considered a precursor of Bohmian mechanics.
These ideas, though, transcend conceptual or philosophical issues,
bringing Bohmian mechanics down to a more applied level, as it has
been the case over the last decade.
During this time, this theory has undergone a rebirth,
passing from being another way to understand quantum mechanics
``without observers''\cite{shelly} to a well-known and increasingly
accepted resource to interpret quantum processes and phenomena, as
well as to devise numerical algorithms to simulate
them.\cite{wyatt-bk,pratim-bk,oriols-bk,sanz-bk}

The main goal of this work consists of helping to develop an
appropriate quantum thinking, i.e., a way to think, to interpret and
to understand quantum mechanics within a more natural scenario than
the standard one, based on abstract and subjective concepts, such
as postulates, operators and probabilities.
When passing from the classical to the quantum-mechanical world, one
usually loses track of concepts which are very much attached to our
daily intuition.
Bohmian mechanics, on the contrary, offers the appropriate tools to
overcome the intermediate gap, allowing us to adapt to the new
conceptual framework by means of already well-known concepts.
In particular, from our own experience, we find quantum concepts
permeate students faster when intuitive models like the Bohmian one are
utilized, for they can move from classical scenarios to
quantum-mechanical ones by means of the same element: trajectories.
As it was already pointed out by Bell,\cite{bell-bk} this constitutes
an important advantage in order to better learn and understand the
physics behind quantum mechanics.

In this regard, it is worth mentioning that the work presented here
can be considered as complementary to the one recently published by
Jeremy Bernstein,\cite{bernstein} ``More about Bohm's quantum''.
In it, Bernstein presents a nice introduction to the Bohmian
interpretation of quantum theory; our goal here is to show how this
theory can be applied to study some well-known quantum processes.
From a pedagogical point of view, these works could be used to cover
the gaps on developments within the quantum theory over the second
half of the twentieth century.\cite{carr}
This is particularly relevant if one considers that the technologies
available at present allow us to explore experimentally scenarios
that were considered only the ground of {\it gedankenexperiments}
until recently, thus reopening the interest in the foundations of
quantum mechanics ---think, for example, about the technological
applications were fundamental processes, such as matter wave
interference or entanglement, are involved.

As it is clear, here we cannot cope with all the issues covered
by quantum mechanics (more detailed accounts can be found in references
like those provided above), so we have focussed
our discussion on the dynamics led by the quantum phase in phenomena
such as quantum coherence, diffraction and interference.
These phenomena have been particularly chosen because of their
relevance in the historical development of quantum mechanics as well
as their important role in different fundamental and technological
applications nowadays.
As it will be shown, when attention is primarily paid to the quantum
probability current density instead of to the probability density,
very interesting and challenging properties arise.
They become very apparent through Bohmian mechanics,
although they could also be found in standard quantum mechanics
with some care, since they are usually ``masked'' (we do not observe
them there, because we rarely look at quantities dynamically
depending on the quantum phase, such as the quantum probability
current density).
This situation is analogous, to some extent, to that of quantum
nonlocality:\cite{bell} although it is always present, it only
emerges strikingly in experiments like the
Einstein-Podolsky-Rosen-Bohm one.\cite{aspect}
Now, far from only being a mere academic exercise, we would also
like to stress the potential interest and power of Bohmian mechanics.
At a fundamental level, due to the insight it provides to analyze all
our preconceived notions of quantum processes and phenomena; at an
applied level, because of the direct implications in fields such as
matter wave interferometry or quantum information, for example.

This work has been organized as follows.
In order to be self-contained, in Section~\ref{sec2}, the basic
elements of Bohmian mechanics are briefly introduced.
In Section~\ref{sec3}, a discussion on the meaning of
the concept of quantum trajectory is presented.
In Sections~\ref{sec4} to \ref{sec6}, we analyze and discuss three
cases of interest where the quantum phase plays a fundamental role
through the properties of quantum coherence and interference.
Finally, the conclusions extracted from this work
are summarized in Section~\ref{sec7}.


\section{Formal grounds of Bohmian mechanics}
\label{sec2}

Within Bohmian mechanics, the wave function
$\Psi$ supplies the quantum system with dynamical information on each
point of the associated configuration space at each time.
This information is encoded in its phase, as can be seen through
the transformation relation
\begin{equation}
 \Psi({\bf r},t) = \rho^{1/2}({\bf r},t) e^{iS({\bf r},t)/\hbar} ,
 \label{e2}
\end{equation}
where $\rho$ and $S$ are the probability density and phase of $\Psi$,
respectively, both being real-valued quantities.
This relation allows us to pass from the time-dependent
Schr\"odinger equation,
\begin{equation}
 i\hbar\ \frac{\partial \Psi}{\partial t} =
  \left( - \frac{\hbar^2}{2m}\ \nabla^2 + V \right) \Psi ,
 \label{e1}
\end{equation}
to the system of coupled equations
\begin{eqnarray}
 \frac{\partial \rho}{\partial t} & + &
  \nabla \cdot \left( \rho\ \frac{\nabla S}{m} \right) = 0 ,
 \label{e3} \\
 \frac{\partial S}{\partial t} & + & \frac{(\nabla S)^2}{2m} +
  V + Q = 0 ,
 \label{e4}
\end{eqnarray}
where
\begin{eqnarray}
 Q \equiv - \frac{\hbar^2}{2m} \frac{\nabla^2 \rho^{1/2}}{\rho^{1/2}}
 = \frac{\hbar^2}{4m}
   \left[ \frac{1}{2} \left( \frac{\nabla \rho}{\rho} \right)^2
   - \frac{\nabla^2 \rho}{\rho} \right]
 \label{e6}
\end{eqnarray}
is the so-called quantum potential.
Equation~(\ref{e3}) is the continuity equation, which describes the
ensemble dynamics, i.e., the motion of a swarm of trajectories
initially distributed according to some $\rho_0$.
Equations~(\ref{e4}) and (\ref{e5}) govern the motion of individual
particles; the quantum Hamilton-Jacobi equation (\ref{e4}) accounts
for the phase-field evolution ruling the quantum particle dynamics
through the equation of motion
\begin{equation}
 {\bf v} = \dot{\bf r} = \frac{\nabla S}{m} .
 \label{e5}
\end{equation}
This relation indicates that one can define a local velocity field on
each point of the system configuration space
[see also Eq.~(\ref{e10}), below] and, by integrating it in time, to
obtain the corresponding trajectory.
In principle, this is a general result that goes beyond any particular
interpretation of the quantum theory (although it is more apparent
with the Bohmian approach) and is totally neglected in standard
quantum mechanics courses.
It stresses in a very nice fashion the important role played
by phase in quantum mechanics (actually, in the dynamics of quantum
systems), which is usually considered only when talking about
interference.

The coupling between Eqs.~(\ref{e3}) and (\ref{e4}) through $Q$ (or,
equivalently, $\rho$ and its space derivatives) is the reason why
quantum (Bohmian) dynamics is very different from its classical
counterpart.
More specifically, in classical mechanics we find
\begin{eqnarray}
 \frac{\partial \rho_{\rm cl}}{\partial t} & + &
  \nabla \cdot \left( \rho_{\rm cl}\ \frac{\nabla S_{\rm cl}}{m}
   \right) = 0 ,
 \label{e3b} \\
 \frac{\partial S_{\rm cl}}{\partial t} & + &
  \frac{(\nabla S_{\rm cl})^2}{2m} + V = 0 ,
 \label{e4b}
\end{eqnarray}
where $S_{\rm cl}$ is the well-known classical action and
$\rho_{\rm cl}$ denotes a classical particle distribution function.
As it can be seen, $S_{\rm cl}$ establishes a coupling between these
two equations, such that the individual particle motion rules the
evolution of the particle distribution.
However, there is no feedback from
the latter on the former, as it happens in the quantum case, where
$\rho$ also governs the individual motion through $Q$.
In other words, this coupling enables the wave function to
guide the particle motion.
Also, it can be noticed that classical trajectories are obtained from
a relation like (\ref{e5}), although $S_{\rm cl}$ is not referred to
as a classical phase, but as the classical action.
This is not just a coincidence, but is related to an analogous concept
to that of quantum phase, used by Schr\"odinger in the derivation of
his equation,\cite{sanz-bk} namely the surfaces of constant action in
the phase space, a linking idea between wave optics and classical
mechanics.

In the literature, we essentially find two types of Bohmian
schemes:\cite{wyatt-bk} analytic and synthetic.
The {\it analytic} scheme is based on first solving Eq.~(\ref{e1})
for $\Psi$ and then substituting the latter into Eq.~(\ref{e5}) to
find the quantum trajectories.
It is usually considered for interpretive purposes.\cite{sanz1}
The {\it synthetic} scheme, based on Eqs.~(\ref{e3}),
(\ref{e4}) and (\ref{e5}) (or their hydrodynamical counterparts), is
aimed at devising computational algorithms to obtain the quantum
trajectories ``on-the-fly'' and then to ``synthesize'' $\Psi$ (or
$\rho$) from them.\cite{lopreore}
This second approach, rooted in Madelung's quantum
hydrodynamics,\cite{madelung} was considered in the 1970s by
Bialynicki-Birula\cite{birula} and Hirschfelder,\cite{hirsch}
although later on Wyatt\cite{lopreore} popularized it through
the so-called {\it quantum trajectory methods}.\cite{wyatt-bk}
In quantum hydrodynamics, the magnitudes of interest are the
probability density, $\rho = R^2 = \Psi^* \Psi$, and the probability
current density, ${\bf J} = \rho {\bf v} = R^2\ (\nabla S/m)$, related
through the continuity equation (\ref{e3}),
\begin{equation}
 \frac{\partial \rho}{\partial t} = - \nabla \cdot {\bf J} .
 \label{e9}
\end{equation}
An equation equivalent to (\ref{e4}) can also be constructed
for ${\bf v}$, giving rise to quantum Euler or Navier-Stokes equations.
Instead of trajectories, the solutions of (\ref{e5}) are better
regarded as fluid streamlines (or flow lines), obtained by integrating
\begin{equation}
 {\bf v} = \dot{\bf r} = \frac{{\bf J}}{\rho} ,
 \label{e10}
\end{equation}
formally equivalent to (\ref{e5}).
These lines follow the flow described by the quantum (probabilistic)
fluid associated with the system.


\section{Streamlines, tracer particles and Bohmian mechanics}
\label{sec3}

Once the formal basis of the theory is settled, a question that
immediately arises is: what is exactly a Bohmian trajectory?
Or, equivalently, what does a Bohmian trajectory actually represent?
Given the connection between Eqs.~(\ref{e1}) and (\ref{e5}), at a formal
level it can be said that a Bohmian trajectory describes the evolution
in time of a specific point of the configuration space associated with
the physical system described by (\ref{e2}).
Actually, analogously to classical mechanics, if the value or state of
a certain system property is specified by some particular configuration
space point, the Bohmian trajectory will show us the time-evolution of
such a state (point).
It is in this sense that Bohmian mechanics can be considered to be more
complete than the standard quantum theory,\cite{bohm} since it goes
beyond the probabilistic description of the latter by allowing us to
monitor the individual evolution of each configuration space point.
In this regard,
the trajectory can therefore represent the path displayed by a particle
in a scattering or an interference experiment, for example, as well as
the evolution of some internal degree of freedom, such as the reaction
coordinate describing the passage from reactants to products in a
chemical reaction.
In either case, we gain an insight that helps us to understand the
process that is going on in a causal manner, i.e., with no need for
appealing to a probabilistic description.

This brings about another important question: are Bohmian trajectories
the real paths followed by the degrees of freedom involved in our
description (regardless of what these degrees of freedom may
represent)?
This question was introduced in the literature
by Scully and coworkers,\cite{scully} who ended up arguing that Bohmian
trajectories are too ``bizarre'' when trying to explain some of the
fundamental experiments of quantum mechanics, like Wheeler's delayed
choice experiment.\cite{wheeler}
Unfortunately, this is only a consequence of the misuse of Bohmian
mechanics, as shown by Hiley and Callaghan,\cite{hiley2} for this
theory just allows us to remove all paradoxes we find in the
quantum theory (``Particles do not go through
both slits at the same time, cats do not end up in contradictory
states such as being simultaneously alive and dead, and there is no
measurement problem.'' [\onlinecite{hiley2}]) and explain quantum
phenomena in a more unambiguous fashion.

Controversies like the one mentioned above have a positive side,
for they lead us to think of the reality of Bohmian trajectories,
i.e., to what extent a quantum particle will follow one of such
trajectories, just as a speck of dust describes a (classical) path
suspended in the air.
For pedagogical purposes, in order to render some light on this issue,
consider the following two well-known classical scenarios.
First, imagine we want to measure some physical quantity of a
classical system (e.g., the position of a pendulum at a given time).
Typically, we do not perform one single measurement, but a number of
them and then we compute their average; the resulting outcome is the
value we eventually assign to such a physical quantity.
Obviously, if the initial state is the same and the quantity can be
measured with infinite accuracy, the same outcome should be obtained
after each measurement, which would be therefore dispersionless.
This means a point in the phase space determined by the
degrees of freedom associated with such a system.
However, under the presence of fluctuations, the outcome will be
described by a certain density distribution function covering a region
of such phase space rather than a point.
This leads us to the theoretical framework of statistical mechanics.
Second, consider a classical fluid.
It consists of many different particles (e.g., atoms, ions, molecules,
etc.), all the degrees of freedom being described by a set of
differential coupled equations, with as many equations as degrees
of freedom are involved, in principle (they can be reduced later on
by means of different constraint conditions, but this is irrelevant
here).
Under some assumptions, one can pass from the microscopic description
of the fluid to a macroscopic one, where equations like the Euler or
Navier-Stokes ones will be rather used.
These equations provide us with a phenomenological description of the
evolution of a continuous fluid without paying any attention to the
particular (microscopic) dynamics of its constituents.
This is essentially the basis of classical hydrodynamics.
Now, to understand the dynamics of such a fluid experimentally and then
to compare it with the theoretical model, one usually proceeds by
sprinkling the fluid with some particles.
These are {\it tracer} particles that help us to visualize the flow
dynamics as they move along the fluid streamlines (the lines along
which the fluid current goes or, equivalently, its energy is
transported).
For example, if we want to observe the evolution of an air stream, we
can use smoke; for a liquid like water, we can make use of another
liquid, like ink, or tinny floating particles, like charcoal dust.
At cosmological scales, hydrodynamical approaches can also be utilized,
considering stars, galaxies or galaxy clusters as tracer particles.

After this discussion, let us go back again to Bohmian mechanics and
the meaning of a quantum trajectory and a Bohmian particle.
In real event-to-event interference experiments,\cite{tonomura,shimizu}
we observe that quantum particles behave as in the first
scenario mentioned above: a single measurement or detection is
meaningless, so we need many of them in order to visualize the pattern
formed and, therefore, to obtain information about either the
diffracted particle or the diffracting object.
Individual real quantum particles thus behave like individual
point-like particles, although their distributions display wave-like
behavior, in accordance with Schr\"odinger's equation (\ref{e1}) or
its Bohmian equivalents, Eqs.~(\ref{e3}) and (\ref{e4}).
Hence, these ensemble properties must be dealt with in terms of
statistical descriptors, namely a density distribution function, with
its role being played in quantum mechanics by the probability density
(or, equivalently, the wave function).
This is in agreement with Born's statistical interpretation of quantum
mechanics.
However, if (individual) particles are regarded as moving along single
trajectories, are these trajectories the ones obtained from (\ref{e5})?
In principle, Bohmian trajectories reproduce all the features of
quantum mechanics and one would be then tempted to think that real
particles always move as Bohmian particles (i.e., obeying a Bohmian
dynamics).
However, if the Bohmian equations are understood as hydrodynamic
equations, the trajectories obtained from (\ref{e5}) should not
be regarded necessarily as the trajectories pursued by real particles,
but rather as the streamlines associated with the quantum
fluid, as in the second classical scenario mentioned above (note that
Schr\"odinger's equation does not usually describe a ``true'' particle,
but a degree of freedom).
That is, in principle, Bohmian particles play the same role as
classical tracer particles, allowing us to infer dynamical properties
of the quantum fluid that appear usually as ``hidden'' when studied by
means of the wave function formalism.


\section{Free Gaussian wave packets}
\label{sec4}

In order to properly understand the dynamical role of the quantum
phase as well as the implications of quantum coherence, consider a
free expanding one-dimensional Gaussian wave packet,
\begin{equation}
 \Psi(x,t) = \left( \frac{1}{2\pi\tilde{\sigma}_t^2} \right)^{1/4}
  e^{-(x - v_0 t)^2/4\tilde{\sigma}_t\sigma_0
  + i p_0 (x - v_0 t)/\hbar + iEt/\hbar} .
 \label{e11}
\end{equation}
This wave packet moves along the classical trajectory $x_{\rm cl} =
v_0 t$ (for simplicity, we assume $x_{{\rm cl},0} = 0$) and its
spreading with times goes as
\begin{equation}
 \sigma_t = |\tilde{\sigma}_t| =
  \sigma_0 \sqrt{1 + \left( \frac{\hbar t}{2m\sigma_0^2} \right)^2} .
 \label{e12}
\end{equation}
Extracting $S$ from (\ref{e11}) and then substituting it into
(\ref{e5}), we find the general expression for the corresponding
quantum trajectories,
\begin{equation}
 x(t) = v_0 t + \frac{\sigma_t}{\sigma_0}\ x_0 ,
 \label{e13}
\end{equation}
with $x_0$ being the trajectory initial position.

In Eq.~(\ref{e13}), we clearly distinguish the two contributions that
rule the behavior with time of a free Gaussian wave packet\cite{sanz-jpa2}
(the same could also be applied to other more general wave packets).
First, there is a classical drift which makes any quantum trajectory to
move alongside the corresponding classical path.
This could be regarded as the classicality criterion in Bohmian
mechanics ---actually, this can be related to Ehrenfest's theorem,
as seen bellow.
The second contribution is a quantum (fluid) drift
associated with the expansion undergone by $\Psi$ with time.
Since the trajectories are ``guided'' by the quantum fluid, the
separation among them will increase at a nonuniform accelerated rate,
\begin{equation}
 \frac{dx}{dt} = \frac{\sigma_0}{\sigma_t} \frac{t}{\tau^2}\ x_0 ,
 \label{e14}
\end{equation}
with $\tau \equiv 2m\sigma_0^2/\hbar$.
In other words, free expansion translates into an accelerated quantum
motion, unlike what happens in classical mechanics.
The classical limit thus consists of keeping this term
relatively small, so that the quantum trajectories evolve parallel
to the classical one, $x_{\rm cl}$, in agreement with Ehrenfest's
theorem (see below).

\begin{figure}
 \begin{center}
  \includegraphics[width=8.5cm]{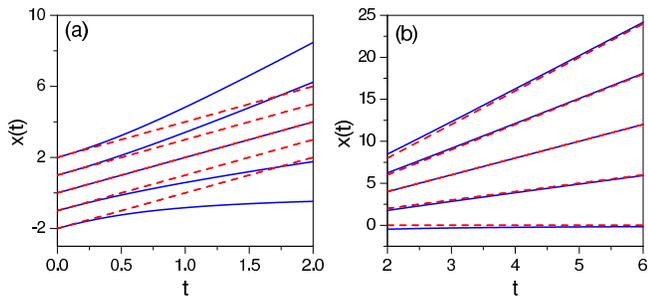}
  \caption{Free evolution of a Gaussian wave packet in terms of
   quantum trajectories (blue solid lines): (a) short-time limit
   (Huygens and Fresnel regimes; the dividing line between both would
   be approximately at $t \approx 0.2$) and (b) long-time limit
   (Fraunhofer regime).
   Some classical counterparts (red dashed lines) undergoing a uniform
   motion are also displayed: in the short-time limit (a), they start
   at $x_{{\rm cl},0} = x_0$ and move with $v_{\rm cl} = v_0$; in the
   long-time limit (b), they start at $x_{{\rm cl},0} = 0$ and move
   with $v_{\rm cl} = v_0 + x_0/\tau$, in agreement with the asymptotic
   expression (\ref{e15}).}
  \label{fig1}
 \end{center}
\end{figure}

By inspecting Eq.~(\ref{e12}), two time regimes become
apparent.\cite{sanz-cpl1}
If $t \ll \tau$, the wave packet spreading is negligible ($\sigma_t
\approx \sigma_0$) and quantum trajectories are all essentially
parallel to $x_{\rm cl}$, since $x(t) \approx x_0 + v_0 t$.
This time scale thus defines a type of classicality criterion that
coincides with the Ehrenfest criterion from traditional courses on
quantum mechanics.\cite{ballentine-bk}
This early stage can be therefore called the Ehrenfest regime, with the
classical drift leading the dynamics.
Since $\Psi$ has not spread too much for these shorter time scales
(i.e., diffraction effects are not relevant yet), it can also be
called the Huygens regime by appealing to an optical analogy
---remember that Huygens' wave theory\cite{bornwolf-bk} does not
account for diffraction phenomena (wave spreading), but only
interference.

As $t$ increases ($t \sim \tau$), we start to observe the action of
the quantum component of the velocity.
This leads to an incipient accelerated motion, which at early stages
can be described according to the familiar expression from classical
mechanics
\begin{equation}
 x(t) \approx x_0 + v_0 t + \frac{1}{2}\ a_q t^2 ,
 \label{e15}
\end{equation}
where $a_q \equiv x_0/\tau^2$ depends on the particle initial position
---the further away this position is with respect to the center of the
wave packet, the faster the particle accelerates.
The result looks as in Fig.~\ref{fig1}(a).
In optics, this is the so-called Fresnel regime, where phases depend
quadratically on coordinates and therefore the wave varies importantly
as it moves small distances from the source.
Similarly, we also find a quantum Fresnel regime as time increases and
the diffraction effects are more noticeable on the wave packet.

As time proceeds and gets relatively large ($t \gg \tau$), the
time-dependence of $\sigma_t$ becomes linear and the wave packet
acquires a stationary shape which does not change with time.
In optics, this is the so-called Fraunhofer regime, where phases
depend linearly on coordinates and therefore the shape of $\Psi$
remains invariant.
Within this regime, quantum trajectories go as
\begin{equation}
 x(t) \approx \left( v_0 + \frac{x_0}{\tau} \right) t ,
 \label{e16}
\end{equation}
i.e., the asymptotic motion is again uniform, but with the
corresponding (constant) velocity having a component proportional to
the initial position of the trajectory, as seen in Fig.~\ref{fig1}(b).
This extra quantum-mechanical component is very important: in grating
diffraction, for example, it is linked to the different diffraction
channels\cite{sanz-talbot} and crystal momentum transfer.\cite{sanz1}

Summarizing, by inspecting the topology displayed by the quantum
trajectories, we note that the quantum flow evolves from an initially
confined fluid to a linearly expanding one, undergoing at times of the
order of $\tau$ a sort of internal boosting which bursts it
open.\cite{sanz-cpl1,hiley3}
If instead of a Gaussian shape, a square one was chosen for the wave
packet, one would observe fractal features when it is
released.\cite{berry}
These features manifest very strikingly when analyzing the
corresponding quantum trajectories, which are also fractal-like
curves.\cite{sanz-jpa1}


\section{Wave-packet superposition}
\label{sec5}

The previous analysis allows us to find practical criteria to discern
whether a coherent superposition of wave packets will display
temporary or permanent interference features in the long-time
dynamics.
To this end, consider the spreading rate $p_s = mv_s \equiv
\hbar/2\sigma_0$, associated with the spreading of the wave
packet.\cite{sanz-jpa2}
Assuming $x_0 \sim \sigma_0$, Eq.~(\ref{e16}) can be expressed as
$x(t) \approx ( v_0 + v_s )\ t$, from which asymptotic interference
properties can be inferred by only considering initial-condition
parameters.
If $v_0 \gg v_s$, a classical-like propagation will be dominant
(the wave packet spreading will not be relevant), while for $v_0 \ll
v_s$ quantum (spreading) effects will appear very rapidly.
Hence, given a wave packet superposition like
\begin{equation}
 \Psi({\bf r},t) = \psi_1({\bf r},t) + \psi_2({\bf r},t) ,
 \label{e19}
\end{equation}
with the $\psi_i = \rho_i^{1/2} e^{iS_i/\hbar}$ being
counter-propagating ($v_{0,2} = - v_{0,1}$) Gaussian wave packets
like (\ref{e11}), two situations are then possible,\cite{sanz-jpa2}
collision-like and interference-like, respectively.
In the first case, the wave packets remain spatially localized after
they interfere; in the latter, they cannot be distinguished
individually due to the permanent presence of (two-slit like)
interference features.

\begin{figure}[t]
 \begin{center}
  \includegraphics[width=8.5cm]{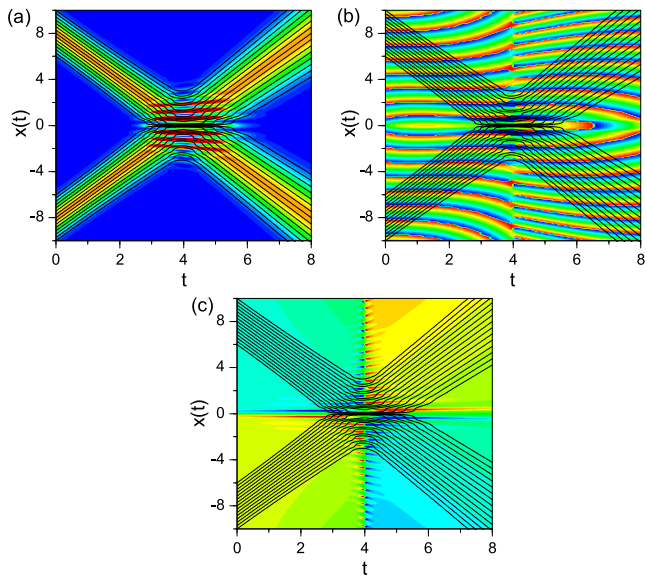}
  \caption{Probability density (a), phase field (b) and velocity field
   (c) for a head-on collision of two coherent Gaussian wave packets
   with $v_0 \gg v_s$.
   The color scale, from blue to red, denotes increasing values of the
   corresponding fields; black solid lines correspond to Bohmian
   trajectories starting at different initial positions.}
  \label{fig2}
 \end{center}
\end{figure}

From Eq.~(\ref{e19}), it is readily seen that
\begin{eqnarray}
 \rho & = & \rho_1 + \rho_2 + 2 \sqrt{\rho_1 \rho_2} \cos \varphi ,
 \label{e20} \\
 {\bf J} & = &
  \frac{1}{m}\ {\rm Re} \left( \Psi^* \hat{\bf p} \Psi \right) =
  - \frac{i\hbar}{2m} \left( \Psi^* \nabla \Psi
    - \Psi \nabla \Psi^* \right) \nonumber \\
  & = & \frac{1}{m} \Big[ \rho_1 \nabla S_1 + \rho_2 \nabla S_2
    + \sqrt{\rho_1 \rho_2} \nabla (S_1 + S_2) \cos \varphi
  \nonumber \\
  & & + \hbar \left( \rho_1^{1/2} \nabla \rho_2^{1/2}
      - \rho_2^{1/2} \nabla \rho_1^{1/2} \right) \sin \varphi \Big] ,
 \label{e21}
\end{eqnarray}
which, when substituted into Eq.~(\ref{e10}), give rise to
\begin{eqnarray}
 \dot{\bf r} & = & \frac{1}{m}
  \frac{\rho_1 \nabla S_1 + \alpha \rho_2 \nabla S_2
  + \sqrt{\alpha} \sqrt{\rho_1 \rho_2} \nabla (S_1 + S_2)
    \cos \varphi}
  {\rho_1 + \alpha \rho_2 + 2 \sqrt{\alpha} \sqrt{\rho_1 \rho_2}
    \sin \varphi}
 \nonumber \\
 & & + \sqrt{\alpha}\ \frac{\hbar}{m}
 \frac{\left( \rho_1^{1/2} \nabla \rho_2^{1/2}
 - \rho_2^{1/2} \nabla \rho_1^{1/2} \right) \sin \varphi}
 {\rho_1 + \alpha \rho_2 + 2 \sqrt{\alpha}
   \sqrt{\rho_1 \rho_2} \cos \varphi} .
 \label{e22}
\end{eqnarray}
Expressions (\ref{e21}) and (\ref{e22}), which can also be derived
in standard quantum mechanics, just contain the essence of this
theory, namely the specific meaning of the concept of quantum coherence.
However, not much attention is usually paid to them, since one usually
focuses on probability densities ---except for calculations of net
fluxes through surfaces, as it is done in tunneling or scattering
problems.
To stress the importance of quantum probability current densities and
quantum velocity fields, consider Eq.~(\ref{e19}) to represent the
head-on collision of two wave packets, as shown in Fig.~\ref{fig2}.
In this figure, the contour plots illustrate how the corresponding
probability density (a), phase field (b) and velocity field (c)
evolve with time; in the three panels, the Bohmian trajectories
(black solid lines) indicate the direction of the quantum flow at each
time.
Although there is no apparent initial overlapping between the wave
packets, the fact that both are present induces very well-defined phase
and velocity fields in the intermediate region, which cannot be
neglected concerning the trajectory or phase dynamics:
this translates into a non-crossing that avoids trajectories coming
from different dynamical regions to coalesce on the same (space)
point at the same time.
This is the so-called non-crossing property of Bohmian
mechanics,\cite{bohm,bohm-hiley-bk,holland-bk} which
has an immediate practical consequence: the phase field has to be
properly implemented in any trajectory-based algorithm aimed at
describing processes with presence of interference in order to
achieve accurate simulations.

\begin{figure}[t]
 \begin{center}
  \includegraphics[width=8.5cm]{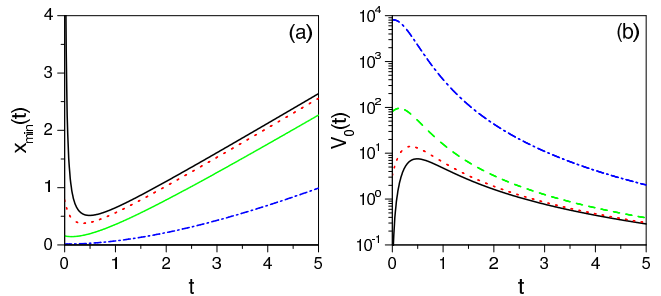}
  \caption{Effective dynamical potential.
   Representation of $x_{\rm min}$ (a) and $V_0$ (b) as a function of
   time for different values of the translational momentum $p_0$.}
  \label{fig3}
 \end{center}
\end{figure}

\begin{figure}[t]
 \begin{center}
  \includegraphics[width=8.5cm]{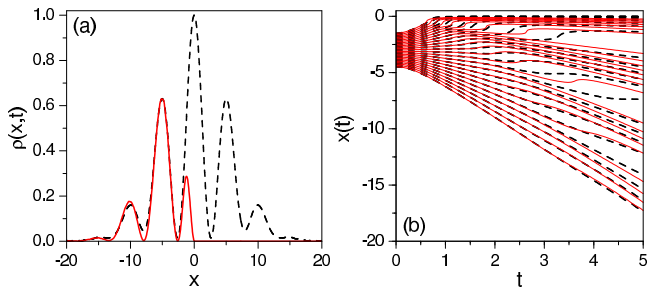}
  \caption{Coherent superposition of two Gaussian wave packets with
   $v_0 \ll v_s$ (black dashed lines):
   (a) probability density at $t=5$ and (b) Bohmian trajectories.
   This problem can be substituted by that of the collision of a single
   Gaussian wave packet with an effective dynamical potential, like
   the one given by Eq.~(\ref{e23}).
   The results associated with the latter problem are displayed with
   red solid line.}
  \label{fig4}
 \end{center}
\end{figure}

The non-crossing property can be interpreted in terms of the presence
of an effective potential that arises from a pure phase effect and,
therefore, is different in nature from the quantum potential $Q$.
As it is known,\cite{hiley4,sanz1,sanz-jpcm} for interference processes
$Q$ presents a rather complex shape initially, in the Fresnel regime;
asymptotically, in the Fraunhofer regime, $Q$
acquires a more regular structure consisting of a series of alternating
plateaus and dips, with the quantum trajectories moving along the
former and avoiding the latter.
The phase-mediated effective potential, on the other hand, allows us to
model the non-crossing effect through the symmetry line splitting the
two regions with opposite (or mirror-symmetric) phase [see
Fig.~\ref{fig2}(b)].
This potential can be simulated by a simple time-dependent
square-well model,\cite{sanz-jpa2}
\begin{equation}
 V(t) = \left\{ \begin{array}{cc}
  0 , & x < x_{\rm min}(t) \\
  - V_0(t) , & \qquad x_{\rm min}(t) \le x \le 0 \\
  \infty , & 0 < x
  \end{array} \right. ,
 \label{e23}
\end{equation}
where the width and depth of the well are
\begin{eqnarray}
 x_{\rm min}(t) & = & \frac{\pi \sigma_t^2}
  {\displaystyle \frac{2p_0\sigma_0^2}{\hbar} +
    \frac{\hbar t}{2m\sigma_0^2}\ x_0} ,
 \label{e24} \\
 V_0(t) & = & \frac{2\hbar^2}{m} \frac{1}{x_{\rm min}^2(t)} ,
 \label{e25}
\end{eqnarray}
respectively.
Thus, while the impenetrable wall gives rise to trajectories bouncing
at $x = 0$, the short-range square well makes the (interference) peak
closer to the wall to have just half the width of the remaining peaks
---in a wave-packet interference pattern every half
of the trajectories that gives rise to the central maximum comes from
dynamical regions with opposite quantum phases or, equivalently,
opposite velocity fields [see Figs.~\ref{fig2}(b)-(c)].
In Fig.~\ref{fig3}, $x_{\rm min}$ (a) and $V_0$ (b) are displayed as a
function of time for different values of the translational momentum
$p_0$.
Thus, in Fig.~\ref{fig4}, it can be noticed how the collision of a wave
packet with this kind of potential reproduces the main features of the
a typical two-slit interference pattern.\cite{sanz-jpa2}
Of course, one could devise more refined models than (\ref{e23}) in
order to get a better agreement with the trajectories coming from the
two interfering wave packets.
However, the important idea to be stressed here is that, from a
dynamical perspective, a problem involving wave packet superposition
is equivalent to the scattering of a single wave packet with an
effective (dynamical) potential ---classically, something similar
happens when two-body collisions are replaced by the collision of a
reduced mass ``effective'' body with an ``effective'' central force.
This highlights the difference between the physics and the mathematics
of the superposition principle.

As seen above, quantum coherence and its Bohmian effect, namely the
non-crossing property, allow us to discern the slit traversed by a
particle without disturbing it in two-slit experiments, at least from
a theoretical point of view.
When dealing with gratings with an infinite number of slits, this
effect leads quantum trajectories to get confined along channels.
That is, trajectories departing from a particular slit keep moving
within the region delimited by the boundaries of such a slit.
This constitutes a physical Bohmian view of the so-called Born-von
Karman boundary conditions for periodic slit gratings or crystal
lattices, which is at the heart of the so-called Talbot
effect,\cite{sanz-talbot} for example.


\section{Application to more complex systems}
\label{sec6}

Here we are going to briefly discuss more complex problems where the
quantum phase and interference dynamics also play an important role
and how Bohmian mechanics may help us to understand them.
Thus, let us start by considering the scattering of rare gas atoms with
metal surfaces with presence of impurities,\cite{sanz-jcp} where
we can observe the appearance of vortical dynamics due to the
overlapping of incoming and outgoing wavefronts.
This mathematical overlapping between two solutions (ingoing and
outgoing waves) gives rise to a physical effect on quantum
trajectories: they display loops and temporary trapping along the
surface.

This type of behavior also plays a fundamental role in problems
involving a bound dynamics, such as those describing the passage from
reactants to products in a chemical reaction.
Here, moreover, the question of tunneling also arises.
As it is known, typical tunneling problems in chemical reactivity
essentially consist of reducing the dimensionality of a complex system
to one (typically, the subspace of the reaction coordinate) and then
compute the corresponding transmission probability by means of some
kinematic or dynamic method.
This can be, however, very misleading.
In one dimension there is only one way to pass from reactants to
products.
Therefore, if there is transmission for an (average) energy below the
top of the barrier (i.e., when it is not possible by means of any
classical mechanism), we speak about tunneling.
In more dimensions, though, this criterion does not have a general
validity, for the passage is ruled by both the total amount of energy
and the orientation of the flow.
Indeed, under certain conditions one can observe\cite{sanz-barna}
that the production of products is more efficient classically than
quantum-mechanically due to the inhibiting effect of interference
(and the associated vortical dynamics).

\begin{figure}
 \begin{center}
  \includegraphics[width=8.5cm]{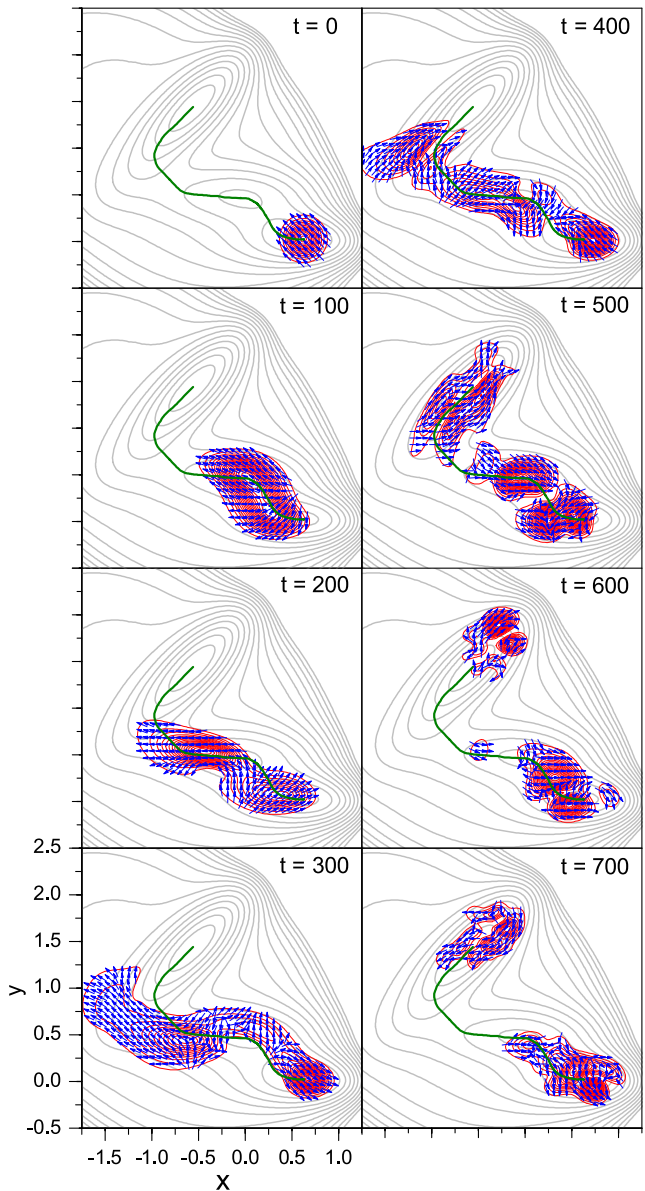}
  \caption{Snapshots illustrating the evolution of the probability
   density contours (red) and velocity-field arrow map (blue) of an
   initial Gaussian wave packet acted by a typical potential (gray
   contours) that describes the passage from reactants to products
   in a chemical reaction.\cite{sanz-barna}
   The reaction path is denoted by the green thicker line.}
  \label{fig5}
 \end{center}
\end{figure}

The reaction dynamics of a typical chemical process is displayed by
means of snapshots in Fig.~\ref{fig5}.
In this figure, the probability density is
represented by a series of contours (red solid lines), the local
direction of the velocity field [as defined by Eq.~(\ref{e10})] is
denoted by the blue arrows and the so-called reaction path joining
reactants with products is denoted with the green thicker line;
the potential energy surface describing the reaction is represented
by the black thinner contours.
In this kind of scenario, it can be shown\cite{sanz-barna} that,
surprisingly, for lower energies the chance of observing formation of
products is larger with classical than with quantum particles.
The reason for this can be found in the formation of ripples that
avoid a larger penetration of quantum trajectories to products, while
a lesser amount of energy classically means that the (classical)
dynamics is smoother and more trajectories can point correctly and
pass to products.
As energy increases, the situation reverts; the pressure of the quantum
fluid is larger and more quantum trajectories can pass to products,
while classically particles have a larger momentum and, therefore, the
dynamics eventually becomes more complicated (chaotic).


\section{Final remarks}
\label{sec7}

The main goal of the present work has consisted of introducing
a comprehensive program that allows a better understanding of the
physical implications of quantum mechanics beyond the standard
formalism.
With such a purpose, the pedagogical advantages of Bohmian mechanics
have been considered, in particular, applying it to quantum coherence,
diffraction and interference.
Thus, by means of Bohmian mechanics, we have seen a
different perspective of quantum phenomena and learnt about an
alternative
information which is usually given by the quantum phase rather than
by density distributions.
The fact that Bohmian mechanics is based on trajectories makes this
information more apparent, since the evolution of Bohmian particles
depend directly on the quantum phase and, therefore, will allows us
to monitor any dynamical feature related with it.

Taking this into account, Bohmian mechanics acquires a potential
interest at both the fundamental and the applied level.
At a fundamental level, it constitutes an ideal framework to analyze
and to rethink all our preconceived notions on quantum processes and
phenomena, i.e., how they take place, as we have seen.
In particular, we have shown how the physics of interference processes
changes dramatically when we switch from a mathematical to a physical
read of the superposition principle.
Although mathematically this principle seems to be a sort of
``innocent'' tool to solve partial differential equations by adding
simpler solutions (something reflected in the own mathematical
structure of Hilbert space), the physical (dynamical) consequences are
quite far from such a simplicity, finding behaviors that go from
non-crossing fluxes to vortical dynamics.
These analyses and concepts can also be straightforwardly extended to
electromagnetic phenomena, which will then be described in a similar
trajectory-like fashion.\cite{sanz-photon}

Obviously, such a fundamental issue has an applied counterpart when
the phenomenon under study is specifically linked to the experiment.
In the case of interference, dealt with here, it has direct
implications in fields such as matter wave interferometry, strong
field ionization, quantum information or quantum control, among others.
In this sense, Bohmian mechanics can be important in the understanding
of processes such as decoherence or quantum erasure.
As we know, the quantum dynamics is closely connected to the quantum
phase and, therefore, any clue on the quantum flux dynamics can be
very important to understand processes where the quantum phase is
relevant.
Note that, except for cases such as the Aharonov-Bohm
effect\cite{aharonov} or the Josephson effect,\cite{josephson} where
the quantum phase plays a key role, not much attention has been paid to
it until the advent of quantum information and quantum control, with
which it has become more relevant.
Even though, the use of quantum current densities and quantum
velocity fields is not so general.
A reason for that could be that, contrary to probability densities (i.e.,
intensities, transmission probabilities, reflection coefficients, cross
sections, etc.), such fields are not observable directly ---although
they give rise to indirect consequences experimentally detectable,
such as the previously mentioned effects.


\vspace{.25cm}
\section*{Acknowledgements}

We are grateful to Prof.\ Basil J.\ Hiley for his valuable criticisms
and careful proofread of the manuscript.
This work has been supported by the Ministerio de Ciencia e
Innovaci\'{o}n (Spain) under Projects FIS2010-18132 and FIS2010-22082,
as well as by the COST Action MP1006 ({\it Fundamental Problems in
Quantum Physics}).
A. S. Sanz would also like to thank the Ministerio de Ciencia e
Innovaci\'{o}n for a ``Ram\'on y Cajal'' Grant.



\begin{thebibliography}{88}
\bibitem{note1}
 Bohmian mechanics is also referred to in the literature as de
 Broglie-Bohm theory, pilot-wave theory, and
 de Broglie-Bohm-Madelung-Takabayasi theory, as well as causal,
 ontological, or hydrodynamical interpretation of quantum mechanics.
 Here, we have considered the first denomination, Bohmian mechanics,
 in order to highligth more explicitly the mechanical explanation it
 provides of quantum phenomena, just like classical mechanics does of
 classical phenomena.
\end{thebibliography}


\end{document}